\documentclass[12pt]{article}

\usepackage{graphicx}
\begin{document}
\begin{center}
{\bf A model of nonlinear electrodynamics }\\
\vspace{5mm}
 S. I. Kruglov
\footnote{E-mail: serguei.krouglov@utoronto.ca}
 \\

\vspace{5mm}
\textit{Department of Chemical and Physical Sciences, University of Toronto,\\
3359 Mississauga Road North, Mississauga, Ontario, Canada L5L 1C6}
\end{center}

\begin{abstract}

A new model of nonlinear electrodynamics with two parameters is investigated. We also consider a model with one dimensional parameter. It was shown that the electric field of a point-like charge is not singular
at the origin and there is the finiteness of the static electric energy of point-like charged particle.
We obtain the canonical and symmetrical Belinfante energy-momentum tensors and dilatation currents. It is
demonstrated that the dilatation symmetry and dual symmetry are broken in the models suggested. We have calculated the static electric energy of point-like particles.

\end{abstract}

\section{Introduction}

A renewal interest in nonlinear electrodynamics is due to the possible quantum gravity corrections to linear electrodynamics. Thus, the new parameter with the dimension of the length is introduced. A non-linear electrodynamics can have a finite electromagnetic energy of a point charge contrarily to Maxwell's electrodynamics. This takes place in the Born-Infeld (BI) electrodynamics \cite{Born}, \cite{Infeld}, \cite{Plebanski}. So called logarithmic electrodynamics and exponential electrodynamics were considered in \cite{Gaete} and \cite{Hendi}, respectively.
The dimensional constant introduced in nonlinear electrodynamics gives the upper bound on the possible electromagnetic fields. Thus, we introduce the parameter $\beta$ with the dimension of length$^{4}$ (or the dimension of mass$^{-4}$).
If the parameter $\beta$ approaches to zero one of the models under consideration converts to Maxwell's
electrodynamics. In this paper, we formulate and investigate two models of nonlinear electrodynamics.
One of the models has the dimensional parameter $\beta$ and the dimensionless parameter $a$. Another model possesses
only one dimensional parameter $\beta$.

The paper is organized as follows. In section 2, the Lagrangians of new models of nonlinear electrodynamics
are introduced. The field equations are formulated which are similar to electrodynamics in the medium. We obtain
the electric permittivity $\varepsilon$ and  magnetic permeability $\mu$ of the vacuum in section 3.
We demonstrate that the electric field of a point-like charge is not singular at the origin and we obtain the maximum of the electric field. In  section 4 the canonical and symmetrical Belinfante energy-momentum tensors, the dilatation
currents, and their non-zero divergences are found. We discuss the results in section 5.

The Heaviside-Lorentz system with $\hbar =c=\varepsilon_0=\mu_0=1$ is used, and Euclidian metric is
chosen so that the 4-vector is $x_\mu=(x_m,ix_0)$, $x_0$ is a time. Greek
letters run from $1$ to $4$ and Latin letters range from $1$ to $3$.

\section{The models}

Let us consider the Lagrangian density of nonlinear electrodynamics
\begin{equation}
{\cal L} = -{\cal F}-\frac{a{\cal F}}{2(\beta{\cal F})+1},
 \label{1}
\end{equation}
where $\beta$ is dimensional parameter so that $\beta{\cal F}$ is dimensionless, and $a$ is a dimensionless parameter. The Lorentz-invariant is ${\cal F}=(1/4)F_{\mu\nu}^2=(\textbf{B}^2-\textbf{E}^2)/2$ and $F_{\mu\nu}=\partial_\mu A_\nu-\partial_\nu A_\mu$ is the field strength ($A_\mu$ is the 4-vector-potential).
It should be noted that Lagrangian (1) is singular at a finite value of the electric field.
We will show that the field of a point charge and the energy contained in it, are finite, and the parameter $\beta$ is connected with the upper bound on the possible field strength. If $a\rightarrow 0$ the Lagrangian density (1) converts to the Maxwell Lagrangian density. Assuming that $\beta{\cal F}\ll 1$, we obtain from Eq.(1) the approximate Lagrangian density
\begin{equation}
{\cal L}\simeq -(1+a){\cal F}+2a\beta {\cal F}^2,
\label{2}
\end{equation}
and we imply that $a\ll 1$.
Thus, at weak electromagnetic fields the Lagrangian density (2) approaches to ${\cal L}\simeq -(1+a){\cal F}$. As a result the proposed model violates the correspondence principle, that requires that in the limit of weak fields the nonlinearity should disappear, and the field equations be the same as the standard Maxwell equations of classical electrodynamics. In the model based on Lagrangian (2) the dielectric permittivity, $\varepsilon$, and magnetic permeability, $\mu$, of the vacuum, $(F_{\mu\nu} =0)$, are different from unity. Note that the correspondence principle is obeyed by QED (see \cite{Heisenberg}, \cite{Schwinger}, \cite{Adler}) and by the BI Lagrangian. One can speculate that there is a contribution of non-electromagnetic fields (for example gravitational fields) to electromagnetic vacuum that results in the appearance of $\varepsilon\neq 0$, $\mu\neq 0$ for ``free" space.

The model with the Lagrangian density containing quadratic terms in Lorentz-invariants was investigated in \cite{Kruglov}. Such nonlinear model of electrodynamics appears due to the vacuum polarization of arbitrary spin particles \cite{Kruglov1}, \cite{Kruglov5}, \cite{Kruglov2}. The model leads to birefringence which means that the phase velocities of light are different for different polarizations of electromagnetic fields.  The model of electromagnetic fields at cubic order in the field strength, violating Lorentz invariance, also leads to birefringence \cite{Kruglov3}. It is known that there is not birefringence in BI electrodynamics. But in generalized BI electrodynamics \cite{Kruglov4} with two constants birefringence takes place.
We also note that the model of ${\cal F}^2$-electrodynamics was considered in the framework
of the Kaluza-Klein theory with the additional Gauss-Bonnet term \cite{Kerner} and also in \cite{Gitman}.
Here we treat the non-linear model under consideration as the effective electrodynamics for strong electromagnetic fields which takes into account, probably, quantum gravity corrections containing fundamental length ($\beta^{1/4}$). The model allows us to consider particular cases for different parameters $a$ and $\beta$.

We also consider the model which does not violate the correspondence principle with the Lagrangian density
\begin{equation}
{\cal L}_1= -\frac{{\cal F}}{2(\beta{\cal F})+1}.
\label{3}
\end{equation}
At small electromagnetic fields, $\beta{\cal F}\ll 1$, the model (3) tends to classical electrodynamics, and the Lagrangian density becomes ${\cal L}\simeq -{\cal F}$. As a result, the correspondence principle is not broken.

\section{The field equations}

From the Euler-Lagrange equations
\[
\partial_\mu\frac{\partial{\cal L}}{\partial(\partial_\mu
A_\nu)}-\frac{\partial{\cal L}}{\partial A_\nu} =0
\]
with the help of Eq.(1), we obtain the equations of motion
\begin{equation}
\partial_\mu\left(F_{\mu\nu}+\frac{aF_{\mu\nu}}{\left[2(\beta{\cal F})+1\right]^2}\right)=0.
\label{4}
\end{equation}
The Lagrangian density (3) leads to field equations
\begin{equation}
\partial_\mu\left(\frac{F_{\mu\nu}}{\left[2(\beta{\cal F})+1\right]^2}\right)=0.
\label{5}
\end{equation}
The electric displacement field is given by
$\textbf{D}=\partial{\cal L}/\partial \textbf{E}$ ($E_j=iF_{j4}$), and we obtain
from Eq.(1)
\begin{equation}
\textbf{D}=\left(1+\frac{a}{\left[2(\beta{\cal F})+1\right]^2}\right)\textbf{E}.
\label{6}
\end{equation}
Thus, the electric permittivity $\varepsilon$ is as follows:
\begin{equation}
\varepsilon=1+\frac{a}{\left[2(\beta{\cal F})+1\right]^2}.
\label{7}
\end{equation}
The magnetic field is defined as $\textbf{H}=-\partial{\cal L}/\partial \textbf{B}$ ($B_j=(1/2)\varepsilon_{jik}F_{ik}$, $\varepsilon_{123}=1$), and we find
\begin{equation}
\textbf{H}= \left(1+\frac{a}{\left[2(\beta{\cal F})+1\right]^2}\right)\textbf{B},
\label{8}
\end{equation}
so that $\textbf{B}=\mu\textbf{H}$, and magnetic permeability is $\mu=\varepsilon^{-1}$.
It follows from Eqs.(7),(8) that for an absence of electromagnetic fields, $F_{\mu\nu}=0$, for the vacuum, $\varepsilon=1+a$, $\mu=1/(1+a)$.
For the model based on (3), one obtains the electric permittivity and magnetic permeability
\begin{equation}
\varepsilon_1=\frac{1}{\left[2(\beta{\cal F})+1\right]^2},~~~~\mu_1=\left[1+2(\beta{\cal F})\right]^2.
\label{9}
\end{equation}

From Eq.(4) we obtain the first pair of Maxwell's equations
\begin{equation}
\nabla\cdot \textbf{D}= 0,~~~~ \frac{\partial\textbf{D}}{\partial
t}-\nabla\times\textbf{H}=0. \label{10}
\end{equation}
The second pair of Maxwell's equation can be found from the Bianchi identity
\begin{equation}
\partial_\mu \widetilde{F}_{\mu\nu}=0,
\label{11}
\end{equation}
($\widetilde{F}_{\mu\nu}$ is a dual tensor) and is given by
\begin{equation}
\nabla\cdot \textbf{B}= 0,~~~~ \frac{\partial\textbf{B}}{\partial
t}+\nabla\times\textbf{E}=0. \label{12}
\end{equation}
Thus, equations (10),(12) represent the nonlinear Maxwell equations because the electric permittivity $\varepsilon$ and magnetic permeability $\mu=\varepsilon^{-1}$ depend on the fields $\textbf{E}$ and $\textbf{B}$.
So, the nonlinear electrodynamics equations (4) with two parameters $a$ and $\beta$ were rewritten in the form of Maxwell's equations (10),(12). The vacuum of such a model acts as a medium with
the electric permittivity $\varepsilon$, (7), and magnetic permeability $\mu=\varepsilon^{-1}$.
We note that the condition for non-birefringence in external magnetic field for most general Lagrangians was formulated in \cite{Shabad}.

From equations (6),(8), we obtain the relation $\textbf{D}\cdot\textbf{H}=\varepsilon^2\textbf{E}\cdot\textbf{B}$, and according to \cite{Gibbons}, the nonlinear electrodynamics under consideration is not the duality symmetrical
because $\textbf{D}\cdot\textbf{H}\neq\textbf{E}\cdot\textbf{B}$. For the model based on the Lagrange density (3) we have the similar equation $\textbf{D}\cdot\textbf{H}=\varepsilon^2_1\textbf{E}\cdot\textbf{B}$.
In BI electrodynamics the dual symmetry is conserved. It should be noted that in
QED due to quantum corrections the dual symmetry is violated. Vacuum birefringence within QED was investigated in \cite{Adler1}, \cite{Biswas}.

In electrostatics, at $\textbf{B}=\textbf{H}=0$, with the point-like source in Eq.(10) we obtain
\begin{equation}
\nabla\cdot \textbf{D}_0=e\delta(\textbf{r})
\label{13}
\end{equation}
having the solution
\begin{equation}
\textbf{D}_0=\frac{e}{4\pi r^3}\textbf{r}.
\label{14}
\end{equation}
Equation (14) using (6) can be written as
\begin{equation}
E_0\left(1+\frac{a}{\left(1-\beta E_0^2\right)^2}\right)=\frac{e}{4\pi r^2}.
\label{15}
\end{equation}
At the origin, $r\rightarrow 0$, Eq.(15) possesses the solution
\begin{equation}
E_0=\frac{1}{\sqrt{\beta}}.
\label{16}
\end{equation}
For the model based on Eq.(3), for the point-like source, Eq.(13) becomes
\begin{equation}
\frac{E_0}{\left(1-\beta E_0^2\right)^2}=\frac{e}{4\pi r^2}
\label{17}
\end{equation}
with the same solution (16) at $r\rightarrow 0$.
The finiteness of the electric field at the origin of the charged particle is due to the singularity of the electric permittivity, and this result associates not only for this special model. The presence of singularity is the general property of all models with finite maximum field. At the same time it was demonstrated in \cite{Gitman} that the presence of maximum field is not necessary for the finiteness of the self-energy of a point charge.
The value (16) represents the maximum field strength which is finite contrary to classical electrodynamics.
The same attractive feature takes place in BI electrodynamics. Introducing unitless variables
\begin{equation}
x=\frac{4\pi r^2}{e\sqrt{\beta}},~~~~y=\sqrt{\beta}E_0,
\label{18}
\end{equation}
Eq.(15) takes the form
\begin{equation}
y^5-\frac{1}{x}y^4-2y^3+\frac{2}{x}y^2+y(1+a)-\frac{1}{x}=0.
\label{19}
\end{equation}
It is questionable to obtain the analytic solution to Eq.(19). At $x\rightarrow 0$ ($r\rightarrow 0$) Eq.(19) reduces to the equation $y^4-2y^2+1=0$ with the solution $y=1$ which is equivalent to Eq.(16). When distance $r$ approaches to infinity, $x\rightarrow \infty$, equation (19) becomes $y^5-2y^3+y(1+a)=0$ having only one real solution $y=0$.
Thus, the electric field of a point-like charge is not singular at the origin,
and we have the finiteness of the electric field in all region $0\leq r< \infty$. From Eq.(19) it is easy to obtain the inverse function $x(y)$:
\begin{equation}
x=\frac{\left(1-y^2\right)^2}{y\left[\left(1-y^2\right)^2+a\right]}.
\label{20}
\end{equation}
The plot of the function $x(y)$ for different values of the parameter $a$ is given in Fig.1.
\begin{figure}[h]
\includegraphics[height=3.0in,width=4.0in]{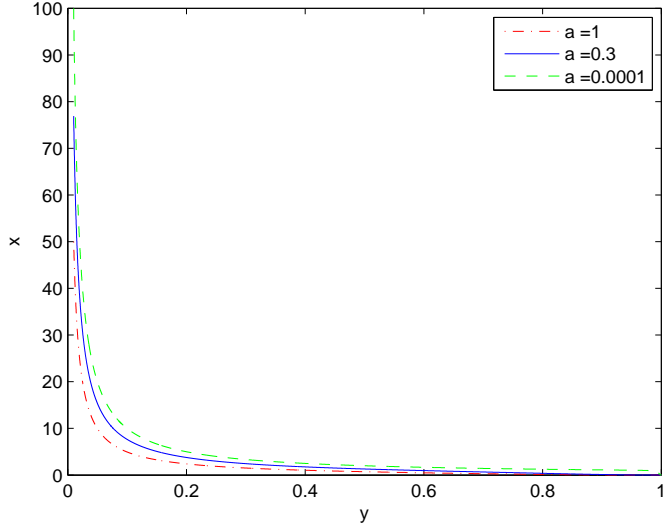}
\caption{\label{fig.1}$x$ versus $y$ for different values of the parameter $a$.}
\end{figure}
The graph shows that, indeed, at $x\rightarrow 0$ the variable $y$ approaches to one, Eq.(16) is valid, and at $x\rightarrow\infty$ we have $y=0$.

Eq.(17) (for the model based on Eq.(3)) in terms of variables (18) becomes
\begin{equation}
y^4-2y^2-xy+1=0.
\label{21}
\end{equation}
At $x\rightarrow 0$, when $r\rightarrow 0$, Eq.(21) possesses the solution $y=1$ leading to Eq.(16).

\section{The energy-momentum tensor and dilatation current}

From the general expression of the canonical energy-momentum tensor
\begin{equation}
T_{\mu\nu}^{c}=(\partial_\nu
A_\alpha)\frac{\partial{\cal L}}{\partial(\partial_\mu
A_\alpha)}-\delta_{\mu\nu}{\cal L},
\label{22}
\end{equation}
with the help of Eq.(1), we obtain
\begin{equation}
T_{\mu\nu}^{c}=-(\partial_\nu A_\alpha)F_{\mu\alpha}\varepsilon-\delta_{\mu\nu}{\cal
L}. \label{23}
\end{equation}
The conservation of the canonical energy-momentum tensor reads $\partial_\mu T^c_{\mu\nu}=0$. The tensor (23) is not
gauge-invariant and symmetrical. One may obtain the symmetric
Belinfante tensor with the aid of the relation \cite{Coleman}:
\begin{equation}
T_{\mu\nu}^{B}=T_{\mu\nu}^{c}+\partial_\beta X_{\beta\mu\nu},
\label{24}
\end{equation}
where
\begin{equation}
X_{\beta\mu\nu}=\frac{1}{2}\left[\Pi_{\beta\sigma}\left(\Sigma_{\mu\nu}\right)_{\sigma\rho}
-\Pi_{\mu\sigma}\left(\Sigma_{\beta\nu}\right)_{\sigma\rho}-
\Pi_{\nu\sigma}\left(\Sigma_{\beta\mu}\right)_{\sigma\rho}\right]A_\rho,
\label{25}
\end{equation}
\begin{equation}
\Pi_{\mu\sigma}=\frac{\partial{\cal L}}{\partial(\partial_\mu
A_\sigma)}=-F_{\mu\sigma}\varepsilon.
\label{26}
\end{equation}
The tensor $X_{\beta\mu\nu}$ is antisymmetrical in first two indexes, $X_{\beta\mu\nu}=-X_{\mu\beta\nu}$, and as a result $\partial_\mu\partial_\beta X_{\beta\mu\nu}=0$. Therefore, $\partial_\mu T^B_{\mu\nu}=\partial_\mu T^c_{\mu\nu}=0$, i.e. the symmetric Belinfante tensor is conserved. The generators of the Lorentz transformations
$\Sigma_{\mu\alpha}$ have the properties
\begin{equation}
\left(\Sigma_{\mu\alpha}\right)_{\sigma\rho}=\delta_{\mu\sigma}\delta_{\alpha\rho}
-\delta_{\alpha\sigma}\delta_{\mu\rho}.
\label{27}
\end{equation}
Using equations (25),(27), we obtain
\begin{equation}
\partial_\beta X_{\beta\mu\nu}=\Pi_{\beta\mu}\partial_\beta A_\nu.
\label{28}
\end{equation}
From equation of motion (4) one finds that $\partial_\mu\Pi_{\mu\nu}=0$. With the help of
Eqs.(26),(28), the Belinfante tensor (24) becomes
\begin{equation}
T_{\mu\nu}^{B}=-F_{\mu\alpha}F_{\nu\alpha}\varepsilon-\delta_{\mu\nu}{\cal L},
\label{29}
\end{equation}
which is symmetrical and conserved tensor. The trace of the energy-momentum tensor (29) is given by
\begin{equation}
T_{\mu\mu}^{B}=\frac{8a\beta {\cal F}^2}{\left[2(\beta{\cal F})+1\right]^2}.
\label{30}
\end{equation}
In linear electrodynamics $a=0$ and the trace of the energy-momentum tensor (30) becomes zero. Non-zero trace (30) appeared in the model under consideration because of the dimensional parameter $\beta$ connected with the fundamental length $l=\beta^{1/4}$. We define, in accordance with \cite{Coleman}, the modified dilatation
current
\begin{equation}
D_{\mu}^{B}=x_\alpha T_{\mu\alpha}^{B}+V_\mu,
\label{31}
\end{equation}
where the field-virial $V_\mu$ is given by
\begin{equation}
V_\mu=\Pi_{\alpha\beta}\left[\delta_{\alpha\mu}\delta_{\beta\rho}
-\left(\Sigma_{\alpha\mu}\right)_{\beta\rho}\right]A_\rho=0,
\label{32}
\end{equation}
and vanishes. Then the modified dilatation current (31) becomes $D_{\mu}^{B}=x_\alpha T_{\mu\alpha}^{B}$,
and the 4-divergence of dilatation current equals
\begin{equation}
\partial_\mu D_{\mu}^{B}=T_{\mu\mu}^B.
\label{33}
\end{equation}
Therefore, the scale (dilatation) symmetry is broken as we have introduced the dimensional parameter $\beta$. The conformal symmetry including the one-parameter dilatation symmetry group is also broken \cite{Coleman} contrarily to linear Maxwell's electrodynamics. We note that in BI electrodynamics the scale symmetry is violated \cite{Kruglov4}.

For the model (3) similar calculations give the Belinfante tensor
\begin{equation}
T_{\mu\nu}^{B}=-F_{\mu\alpha}F_{\nu\alpha}\varepsilon_1-\delta_{\mu\nu}{\cal L}_1,
\label{34}
\end{equation}
were $\varepsilon_1$ is given by Eq.(9) and ${\cal L}_1$ is defined by Eq.(3).
The trace of the energy-momentum tensor (34) becomes
\begin{equation}
T_{\mu\mu}^{B}=\frac{8\beta {\cal F}^2}{\left[2(\beta{\cal F})+1\right]^2}.
\label{35}
\end{equation}
The 4-divergence of dilatation current is given by Eq.(33).

\subsection{Energy of the point-like charge}

From Eq.(29) we obtain the energy density
\begin{equation}
\rho=T^B_{44}=E^2\left(\varepsilon-\frac{1}{2}\right)+\frac{B^2}{2}+\frac{a\left(B^2-E^2\right)}{2\beta
\left(B^2-E^2\right)+2}.
\label{36}
\end{equation}
From Eq.(36), for pure electric energy density, we have
\begin{equation}
\rho_E=E^2\left(\varepsilon-\frac{1}{2}\right)-\frac{aE^2}{2-2\beta
E^2}.
\label{37}
\end{equation}
In terms of unitless variables (18) the normalized total energy $\beta^{1/4}{\cal E}$ in all space, ${\cal E}=\int \rho_E dV$, becomes
\begin{equation}
\beta^{1/4}{\cal E}=\frac{e^{3/2}}{8\sqrt{\pi}}\int_0^\infty y^2\left[1+\frac{a(1+y^2)}{(1-y^2)^2}\right]\sqrt{x}dx,
\label{38}
\end{equation}
where the function $y(x)$ is given by the real root of equation (19). Taking into account Eq.(20), one can represent (38) as
\begin{equation}
\beta^{1/4}{\cal E}=\frac{e^{3/2}}{8\sqrt{\pi}}\int_1^0 \frac{\left[\left(y^2-1\right)^3-a\left(1+3y^2\right)\right]\left[(1-y^2)^2+a\left(1+y^2\right)\right]}
{\sqrt{y}\left[\left(1-y^2\right)^2+a\right]^{5/2}}dy.
\label{39}
\end{equation}
According to Eq.(39), the values of $\beta^{1/4}{\cal E}$ for different parameters $a$ are given in the Table 1.
\begin{table}[ht]
\caption{Normalized total energy $\beta^{1/4}{\cal E}$ of a point-like particle }
\vspace{5mm}
\centering
\begin{tabular}{c c c c c c c c c c c c}
\hline \hline 
$a$ & $0.001$ & $0.01$ & $0.1$ & $0.2$ & $0.3$ & $0.4$ & $0.5$ \\[0.5ex] 
\hline 
$\beta^{1/4}{\cal E}$ & $0.0624$ & $0.0610$ & $0.0556$ & 0.0519 & 0.0490 & 0.0467 & 0.0447\\
[1ex] 
\hline 
\end{tabular}
\label{table:crit}
\end{table}
There was a suggestion that the electron has a pure electromagnetic nature and the mass of the electron equals the finite energy of the field \cite{Born1}, \cite{Rohrlich}, \cite{Spohn}.
If one takes the energy of a point-like charge to be equal the mass of the electron, ${\cal E}=0.51$~MeV, for a parameter $a=0.01$, we obtain the length $l=\beta^{1/4}\approx 23.6$ fm. One can speculate that the model under consideration allows us to realize the old idea of Abraham and Lorentz about the electromagnetic nature of the electron.

For the second model with the Lagrangian density (3), we obtain from Eq.(34) the electric energy density
\begin{equation}
\rho_{1E}=\frac{E^2(1+\beta E^2)}{2(1-\beta E^2)^2}.
\label{40}
\end{equation}
With the help of Eq.(40) the normalized total energy $\beta^{1/4}{\cal E}_1$ in all space is given by
\begin{equation}
\beta^{1/4}{\cal E}_1=\frac{e^{3/2}}{8\sqrt{\pi}}\int_1^0 \frac{(1+y^2)\left(3y^4-2y^2-1\right)}
{\sqrt{y}\left(1-y^2\right)}dy\approx 0.0503.
\label{41}
\end{equation}
Putting ${\cal E}=0.51$~MeV, one obtains the value $l_1=\beta^{1/4}\approx19.48$ fm.

\section{Conclusion}

We have formulated new models of nonlinear electrodynamics. One of the models contains two
parameters $a$ and $\beta$ and the second model has only one parameter $\beta$. The model with the parameter $a$ violates the corresponding principal but allows us to consider the vacuum which is not ``empty" having the dielectric permittivity $\varepsilon=1+a$, and magnetic permeability $\mu=1/(1+a)$. The unitless parameter $a$ characterizes the amount on non-linearity contributed to linear electrodynamics. The second model possessing only one parameter $\beta$ does not violate the corresponding principal and has the standard electromagnetic vacuum with $\varepsilon=1$ and $\mu=1$. We can treat the value $l\equiv\beta^{1/4}$ as a fundamental length due to quantum gravity effects. It was demonstrated that the electric field of a point-like charge
is regular at the origin, and the maximum possible electric field equals $E_{max}=\beta^{-1/2}$ for both models.
It was shown that there is the finiteness of the electromagnetic energy for models under consideration.
We have obtained the canonical and symmetrical Belinfante energy-momentum tensors, and the dilatation current. The dilatation current is not conserved due to the non-zero trace of the energy-momentum tensor in the models under consideration. The scale symmetry is violated because of the dimensional parameter $\beta$ and the dual
symmetry is also broken. The finite static electric self-energy of point-like particles was calculated showing the possibility to treat the value of the mass of the electron as a pure electromagnetic energy.

\end{document}